\documentclass[prl,twocolumn,aps]{revtex4}

\usepackage{graphicx} 

\begin{document}

\title{Towards practical classical processing for the surface code}

\author{Austin G. Fowler, Adam C. Whiteside, Lloyd C. L. Hollenberg}
\affiliation{Centre for Quantum Computation and Communication
Technology, School of Physics, The University of Melbourne, Victoria
3010, Australia}

\date{\today}

\begin{abstract}
The surface code is unarguably the leading quantum error correction
code for 2-D nearest neighbor architectures, featuring a high threshold error rate of approximately 1\%,
low overhead implementations of the entire Clifford group, and
flexible, arbitrarily long-range logical gates. These highly desirable features come at the cost of
significant classical processing complexity. We show how to perform
the processing associated with an $n\times n$ lattice of qubits,
each being manipulated in a realistic, fault-tolerant manner, in
$O(n^2)$ average time per round of error correction. We also
describe how to parallelize the algorithm to achieve $O(1)$ average
processing per round, using only constant computing resources per
unit area and local communication. Both of these complexities are
optimal.
\end{abstract}

\maketitle

Quantum computing promises exponentially faster processing of
certain problems, including factoring \cite{Shor94b} and simulating
quantum physics \cite{Lloy96}. Many quantum algorithms are now known
\cite{Jord10}. The primary challenges are to mitigate and cope with
the imperfections of quantum devices. The surface code
\cite{Brav98,Denn02} supports a powerful quantum computing scheme
\cite{Raus07,Raus07d,Fowl08} featuring an experimentally realistic
threshold error rate of approximately 1\% \cite{Wang11} and
requiring only a 2-D square lattice of qubits with nearest neighbor
interactions. In this work, we describe how to perform the complex
classical processing associated with the full, fault-tolerant scheme
in a complexity-optimal manner. Using the current version of our
code, we can simulate the fault-tolerant operation of millions of
qubits, four orders of magnitude more than in any previous work. A detailed timing analysis can be found in \cite{Fowl12c}.

Previous works on the classical processing of topological quantum
error correction (QEC) have obtained results by making one of two
significant modifications to the problem. Large lattice sizes have
been simulated \cite{Ducl09}, however only by assuming that 4-qubit
operators can be measured perfectly. Small lattices have been
simulated fault-tolerantly \cite{Raus07d,Barr10,Wang09,Wang11},
however these works used the code of Kolmogorov \cite{Kolm09} which does
not support continuous processing of an arbitrary number of rounds
of QEC. Our code now supports continuous fault-tolerant processing,
using constant memory and with processing rate independent of the
number of rounds.

The surface code involves a 2-D lattice of qubits with local
stabilizers \cite{Brav98}. We shall initially assume perfect stabilizer measurement to compare with \cite{Ducl09}.
Error chain endpoints anticommute with stabilizers leading to -1 stabilizer measurements. Each -1 is associated with a vertex. Assuming independent errors, long error chains are exponentially unlikely. Edges between vertices $v_1=(i_1,j_1)$, $v_2=(i_2,j_2)$ are given a weight $w_{12}$ to equal to their Manhattan separation $|i_2-i_1|+|j_2-j_1|$. Edges to boundaries are given a weight equal to their length.

\begin{figure}
\begin{center}
\resizebox{85mm}{!}{\includegraphics{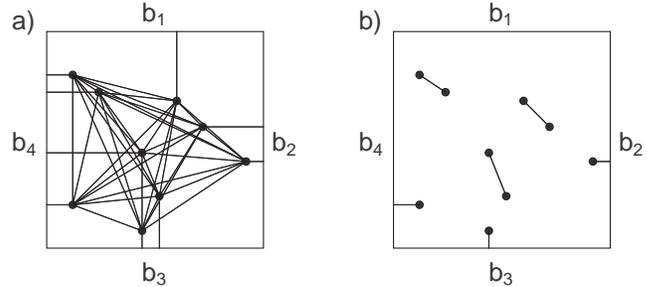}}
\end{center}
\vspace{-5mm}\caption{a) Example of a weighted graph with edges
connecting vertices to labeled boundaries $b_i$. b) Set of edges
such that every vertex is incident on exactly one edge and the total
weight of all edges is minimal.}\label{2Dmatching}
\end{figure}

We independently correct $X$ and $Z$ errors using the minimum weight
perfect matching algorithm \cite{Edmo65a,Edmo65b}. We use our own
implementation which includes the concept of
boundaries and returns edges such that every vertex is
incident on exactly one edge and the total weight is
minimal (Fig.~\ref{2Dmatching}). Corrections are applied along the chosen edges. Logical errors occur when after correction a chain of errors connecting opposing boundaries remains. Provided error chains are well separated, this is unlikely.

\begin{figure}
\begin{center}
\resizebox{85mm}{!}{\includegraphics{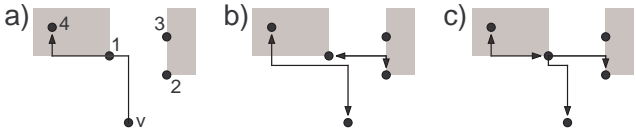}}
\end{center}
\vspace{-5mm}\caption{a) Plane as seen from vertex $v$. Vertices 1
and 2 are unshadowed, vertex 3 is shadowed, vertex 4 is deeply
shadowed. b) A matching with an edge to a deeply shadowed vertex. c)
An equal weight rearrangement of the matching. A lower weight
matching is possible.}\label{shadowing}
\end{figure}

Clearly, a complete graph cannot be used if an $O(n^2)$ runtime is
to be achieved as a complete graph has $O(n^4)$ edges. Simply generating a complete graph would therefore take a minimum of $O(n^4)$ time. Progress can
be made by noting that we are initially only considering data qubit
errors and that a chain of errors between two -1 stabilizer
measurements has length given by the Manhattan metric. If we choose
a vertex and imagine looking at the graph from that vertex, nearby
vertices casts shadows, where we define a point in the plane to be
in shadow if it can be reached by a minimum length path passing
through another vertex (Fig.~\ref{shadowing}a). We define a vertex
to be shadowed if it is in shadow yet neighbors an unshadowed point,
and deeply shadowed if all neighboring points are shadowed. Empirically, we find that if two
vertices are deeply shadowed when viewed from one another, there
always exists a minimum weight perfect matching that does not use an
edge between them (Figs.~\ref{shadowing}b--c). This implies that
such edges do not need to be included in the graph that describes
the problem. This cuts the number of edges down to
$O(n^2)$. The validity of this approach has been verified with
millions of simulations finding identical weight matchings with both
the complete and shadowed approaches.

\begin{figure}
\begin{center}
\resizebox{85mm}{!}{\includegraphics{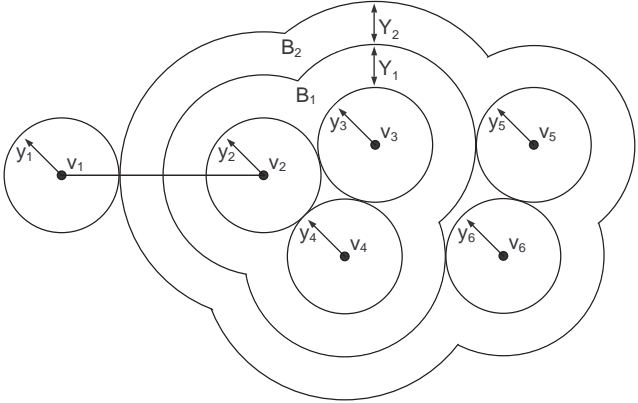}}
\end{center}
\vspace{-5mm}\caption{An example of a tight edge. Edge $e_{12}$ has
the property that $w_{12}-y_1-y_2-Y_1-Y_2=0$.}\label{tight_edge}
\end{figure}

We now describe the minimum weight perfect matching algorithm,
focusing on what is actually done. The original papers of Edmonds
\cite{Edmo65a,Edmo65b} beautifully prove that the method works.
First, some definitions. Let $G$ be a graph with vertices $\{v_i\}$,
edges $\{e_{ij}\}$, and edge weights $\{w_{ij}\}$. Associate with
each vertex $v_i$ a variable $y_i$, which can be thought of as the
radius of a ball centered at $v_i$. Odd sets of vertices can also be
made into blossoms $B_k$ that have their own variables $Y_k$, which
can be thought of as the width of shell around every object in
$B_k$. If a pair of blossoms intersect, then one is contained in the
other. Define an edge $e_{ij}$ to be tight if $w_{ij}-y_i-y_j-\sum
Y_k = 0$, where the sum is over $k$ such that exclusively $v_i$ or
$v_j$ is in $B_k$. This condition is pictorially depicted in
Fig.~\ref{tight_edge}.

\begin{figure}
\begin{center}
\resizebox{85mm}{!}{\includegraphics{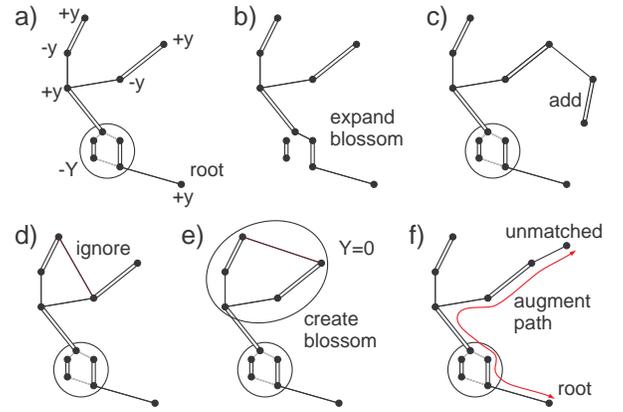}}
\end{center}
\vspace{-5mm}\caption{All required alternating tree manipulations.
a)~Increase outer node and decrease inner node $y$ values (or $Y$ if the node is a blossom), maintaining the tightness of all tree edges and potentially creating new tight edges connected to at least one outer node. b)~Inner blossoms
with $Y=0$ can be expanded into multiple inner and outer nodes and potentially some nodes that are no longer part of the tree. c)~Outer--matched tight edges can be used to grow the alternating tree.
d)~Outer--inner tight edges can be ignored. e)~Outer--outer tight edges
make cycles that can be used to make blossoms. f)~When another unmatched vertex $v$ is found, or an edge to a boundary $b$, the path from the unmatched vertex within the root node through the alternating tree to $v$ or $b$ is augmented, meaning matched edges become unmatched and unmatched edges become matched. This strictly increases the total number of vertices.}\label{Tree}
\end{figure}

Define a node to be a vertex or blossom. Define a blossom to be
unmatched if it contains an unmatched vertex. An alternating tree is
a tree of nodes rooted on an unmatched node such that every path
from the root to a leaf consists of alternating unmatched and
matched edges. Alternating trees can only branch from the root and
every second node from the root. Define branching nodes to be outer.
Define all other nodes in the alternating tree to be inner.
Fig.~\ref{Tree} shows all necessary alternating tree manipulations.

Given a weighted graph $G$, the following algorithm finds a minimum
weight perfect matching.
\begin{enumerate}
\vspace{-2mm}\item If there are no unmatched vertices, return the list of matched edges.

\vspace{-2mm}\item Choose an unmatched vertex $v$ to be the root of alternating tree.

\vspace{-2mm}\item If no edges emanating from the outer nodes of the
alternating tree are tight, henceforth called $O$-tight edges,
increase the value of $y$ or $Y$ associated with each outer node
while simultaneously decreasing the value of $y$ or $Y$ associated
with each inner node until an edge becomes $O$-tight, or an inner
blossom node $Y$ variable becomes 0 (Fig.~\ref{Tree}a).

\vspace{-2mm}\item If an inner blossom node $Y$ variable becomes 0
and there are still no $O$-tight edges, expand that blossom and
return to 3 (Fig.~\ref{Tree}b).

\vspace{-2mm}\item Choose an $O$-tight edge $e$.

\vspace{-2mm}\item If $e$ leads to a matched node not already in the
alternating tree, add the relevant unmatched and matched edge and
associated nodes to the alternating tree and return to 3
(Fig.~\ref{Tree}c).

\vspace{-2mm}\item If $e$ leads to an inner node, mark $e$ so it is
not considered again during the growth of this alternating tree and
return to 3 (Fig.~\ref{Tree}d).

\vspace{-2mm}\item If $e$ leads to an outer node, add the unmatched
edge to the alternating tree. There will now be a cycle of odd
length. Collapse this cycle into a new blossom and associate a new
variable $Y=0$ (Fig.~\ref{Tree}e). Return to 3.

\vspace{-2mm}\item If $e$ leads to an unmatched vertex or boundary,
add $e$ to the alternating tree and augment the path
(unmatched$\leftrightarrow$matched) from the unmatched vertex within the root node to the end of $e$
(Fig.~\ref{Tree}f). Destroy the alternating tree, keeping any newly
formed blossoms. Return to 1.
\end{enumerate}

On average, the algorithm only needs to consider a small local
region around each vertex to find another unmatched vertex to pair
with. This is a property of the graphs associated with topological
QEC only, as the probability of needing to consider an edge of
length $l$ decreases exponentially with $l$. This ensures that the
runtime is $O(n^2)$.

\begin{figure}
\begin{center}
\resizebox{85mm}{60mm}{\includegraphics{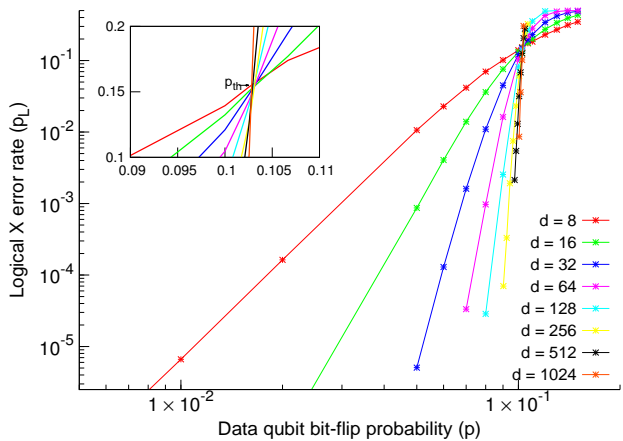}}
\end{center}
\vspace{-5mm}\caption{Logical $X$ error rate $p_L$ versus data qubit
bit-flip probability $p$ for various code distances $d$ assuming
perfect stabilizer measurement. The threshold error rate is
10.25\%.}\label{2-D}
\end{figure}

If we consider a standard square surface code with smooth boundaries
top and bottom and rough boundaries left and right \cite{Fowl08}, we
can randomly apply bit-flips $X$ with probability $p$ to the data
qubits, perfectly measure the $Z$-stabilizers, construct a shadowed
graph as described above, perform minimum weight perfect matching,
apply corrections along the matched edges, and test for logical
failure by checking whether there are an odd or even number of
bit-flips along the top boundary. After correction, there can only
be an odd number of bit-flips along the top boundary if a chain of
bit-flips has formed from top to bottom boundary, indicating a
logical error. The shortest topologically nontrivial chain is the
distance $d$ of the code ($n=2d-1$). By performing many simulations, the
probability of logical $X$ error $p_L$ versus $p$ can be plotted for
a variety of distances (Fig.~\ref{2-D}).

We observe a threshold error rate of 10.25\%, versus 8.2\% in the
work of \cite{Ducl09}, only slightly below the known ideal threshold
error rate of 10.9\% obtained using computationally inefficient
techniques \cite{Ohze09,deQu09}. Furthermore, by comparing the
fraction of the threshold error rate at which our curves cut a
logical error rate of $4\times 10^{-3}$ to the equivalent quantities
in \cite{Ducl09}, it can be seen that our approach corrects twice as
many errors at a given code distance, leading to a quadratic
improvement of the logical error rate. Finally, the raw speed of our
code is evident in our ability obtain data far below threshold.

High performance when assuming perfect stabilizer measurements
enables comparison with existing results, however this case is not
particularly interesting in practice. Only a fully fault-tolerant
approach can be considered practical. When using fault-tolerant
circuits to measure stabilizers and applying depolarizing noise, the
simple 2-D square lattice with Manhattan metric considered above
becomes a complicated 3-D lattice \cite{Wang11} that has both
spatial boundaries and a temporal boundary representing the latest
round of stabilizer measurements. A number of modifications to the
algorithm are required to account for these differences.

Firstly, calculating the complete shadow of a vertex does not work
very well in the 3-D lattice. There are 12 outward directions to
consider instead of 4 \cite{Wang11}, and the probability that all 12
directions are blocked by nearby vertices is low. Instead of
exploring the entire unshadowed region, which can be exceedingly
large, one needs to set a maximum radius of initial exploration.
With high probability, only this initial region is required. Regions
further from the vertex are only explored as required (if another
unmatched vertex or boundary is not found in the initial region).
The probability of requiring a region of radius $r$ decreases
exponentially with $r$. In practice, we choose $r$ just large enough
to ensure nearest and next nearest neighbor lattice locations are
explored initially.

Secondly, the mobile temporal boundary introduces additional
complexity. One must add new vertices to the problem as new data is
obtained. This is straightforward, essentially just increasing the
list of unmatched vertices. However, when growing an alternating
tree, it is possible for the tree to attempt to grow into the
future. We chose to solve this problem by undoing the growth of the
tree and all of the changes its growth introduced --- essentially
running the algorithm backwards. Growth is only reattempted when
data from more rounds of error correction is available.

Thirdly, detecting logical errors is more complex. One needs turn
off errors, perform a perfect round of stabilizer measurement, match
all vertices, apply corrections, detect logical errors as before by
considering the top boundary, and then undo everything to return the
simulator to its state before logical error detection. Note that
this would not be done in a real computer. This procedure consumes
negligible additional memory as a do/undo approach is used, and
takes approximately 10 times as long as a single round of standard
error correction.

Fourthly, it is nontrivial to ensure the amount of memory used
remains finite as the simulation proceeds. With vanishingly small
probability, in principle the entire history of stabilizer
measurements is required to correctly match current time data. In
practice, we keep track of the maximum distance in the past the
algorithm has traversed to perform any matching, and store a fixed
multiple of this distance, discarding any older data. Close to the
threshold error rate, and especially above threshold, extremely
large blossoms are created, involving thousands of vertices and
hundreds of blossom layers. Storing these blossoms dominates the
memory cost. The growth of large blossoms, which is a gradual process, provides a warning that
more data needs to be stored, ensuring reliable processing.

\begin{figure}
\begin{center}
\resizebox{85mm}{60mm}{\includegraphics{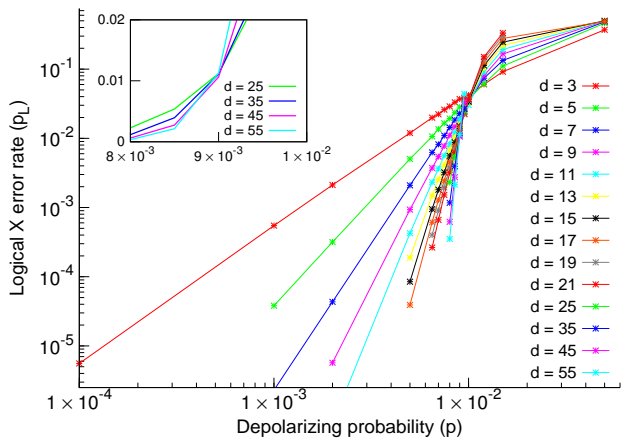}}
\end{center}
\vspace{-5mm}\caption{Logical $X$ error rate per round of error
correction $p_L$ versus depolarizing probability $p$ for various
code distances $d$ using fault-tolerant stabilizer measurement. The
threshold error rate is 0.9\%.}\label{3-D}
\end{figure}

With these modifications in mind, fault-tolerant simulation data is
shown in Fig.~\ref{3-D}. All data points represent one or more
simulation instances run continuously until a total of 10,000
logical errors were observed, enabling reliable determination of the
probability of logical error per round of error correction. Note
that the threshold error rate is 0.9\%, in contrast to prior work
that estimated it at 1.1\% \cite{Wang11}. This highlights the danger
of estimating thresholds from small distance data only. The new code
exactly reproduces the curves of the old code up to $d=13$, however
the true threshold error rate is really only visible for $d>21$, distances at which the old code can not be run.
This lowered threshold can be attributed to boundary effects, since boundary
stabilizers are lower weight and hence more reliable. The physical
error rates at which a factor of 2 (10) improvement in logical error
rate is observed when the code distance is increased by 2 remain
0.5\% (0.2\%) respectively, so this threshold error rate change is
of no practical significance.

Memory limitations prevented the gathering of statistics near the
threshold error rate at higher distances, however at $p=10^{-3}$ the
vast blossoms that make high error rate correction so difficult do
not occur, and it is straightforward to simulate distances as high
as $d=1000$ --- over 4 million qubits. Needless to say, no logical
errors are observed in such a simulation. Each round of error
correction takes under 3 seconds using a single core of an AMD
Opteron, with each individual matching taking tens of microseconds.
Given the algorithm uses only local information, it is in principle
straightforward to parallelize, using a 2-D array of processors with
each processor handling a fixed size patch of code. Inserting a
small pause after each patch correction enables any patch with a
randomly harder matching to catch up after lagging behind. This
would enable the classical processing of an infinite lattice in
constant average time per round of error correction, with the
average time approaching the time for a single matching in the limit
of low $p$ and high classical computing resources.

In summary, we have introduced an algorithm that finds a minimum
weight perfect matching in $O(n^2)$ time given a graph generated by
an $n\times n$ lattice of qubits running the surface code
fault-tolerantly. This algorithm parallelizes to $O(1)$ on an
infinite lattice with constant computing resources per unit area. It
is conceivable that a parallel implementation could achieve hundred
microsecond processing of a round of error correction, sufficient to
keep pace with ion trap quantum gates \cite{Hann09}. Additional
ideas, including implementation in hardware, would be required to
achieve the submicrosecond processing times required to keep pace
with faster gates such as those found in superconducting circuits
\cite{Mari11}.

We acknowledge support from the Australian Research Council Centre
of Excellence for Quantum Computation and Communication Technology
(Project number CE110001027), and the US National Security Agency
(NSA) and the Army Research Office (ARO) under contract number
W911NF-08-1-0527.

\bibliography{../References}

\end{document}